\documentclass[a4paper]{llncs}

\usepackage{color}
\usepackage{url}
\usepackage{graphicx}
\usepackage{tabularx}
\usepackage{multirow,booktabs}
\usepackage{hhline}
\usepackage{times}
\usepackage{arydshln}
\usepackage[square,comma,numbers,sort&compress,sectionbib]{natbib}
\usepackage{mathtools}
\usepackage{paralist}
\usepackage{enumitem}
\setlist{nosep}
\usepackage[usenames,dvipsnames]{xcolor}
\usepackage{microtype}
\usepackage{mathtools}
\usepackage{amsmath,amssymb,latexsym}
\usepackage{wrapfig}
 \usepackage{setspace}
 
\usepackage[skip=0pt]{caption}
\DeclarePairedDelimiter\floor{\lfloor}{\rfloor}

\makeatletter

\renewcommand\section{\@startsection{section}{1}{\z@}{-10\p@ \@plus -4\p@ \@minus -4\p@}{2\p@ \@plus 4\p@ \@minus 4\p@}{\normalfont\large\bfseries\boldmath\rightskip=\z@ \@plus 8em\pretolerance=10000 }}
 
\newcommand\OurSection{\@startsection{subsection}{2}{\z@}{-2\p@ \@plus -4\p@ \@minus -4\p@}{-0.5em \@plus -0.22em \@minus -0.1em}{\normalfont\normalsize\bfseries}}

\makeatother 

\newcommand{\OurMethod}{HiTR}

\usepackage[bookmarks=true%
,bookmarksnumbered=true%
,hypertexnames=false%
,breaklinks=true%
,colorlinks=true%
	,linkcolor=darkblue
	,urlcolor=darkblue
	,anchorcolor=darkblue
	,citecolor=darkblue
]{hyperref}

\definecolor{darkblue}{rgb}{0.0,0.0,0.5}

\newcommand{\MM}[1]{\textcolor{orange}{MM: #1}}

\newcommand{\conditional}{\mspace{2mu} | \mspace{2mu}}
\newcolumntype{C}{>{\centering\arraybackslash}X}

\begin{document}

\mainmatter

\title{Measuring Topical Diversity of Text Documents\\ Using Hierarchical Parsimonization}
\title{Hierarchical Re-estimation of Topic Models for Measuring Topical Diversity}

\author{Hosein Azarbonyad
\and Mostafa Dehghani 
\and Tom Kenter
\and Maarten Marx
\and Jaap Kamps
\and Maarten de Rijke}
\institute{University of Amsterdam, Amsterdam, The Netherlands
\email{\{h.azarbonyad,dehghani,tom.kenter,maartenmarx,kamps,derijke\}@uva.nl}}
\maketitle
\begin{abstract}

A high degree of topical diversity is often considered to be an important characteristic of interesting text documents. 
A recent proposal 
for measuring topical diversity identifies three elements for assessing diversity: words, topics, and documents as collections of words.
Topic models play a central role in this approach.
Using standard topic models for measuring diversity of documents is suboptimal due to \emph{generality} and \emph{impurity}.
General topics only include common information from a background corpus and are assigned to most of the documents in the collection.
Impure topics contain words that are not related to the topic; impurity lowers the interpretability of topic models and impure topics are likely to get assigned to documents erroneously. 
We propose a hierarchical re-estimation approach for topic models to combat generality and impurity;
the proposed approach operates at three levels: words, topics, and documents.
Our re-estimation approach for measuring documents' topical diversity outperforms the  state of the art on PubMed dataset  which is commonly used for diversity experiments.
\end{abstract}

\newcommand{\rqone}{How effective is \OurMethod{} in measuring topical diversity of documents? How does it compare to the state-of-the-art in addressing the general and impure topics problem?}
\newcommand{\rqthree}{Does TR increase the purity of topics? If so, how does using the more pure topics influence the performance in topical diversity task?}
\newcommand{\rqfour}{How does TAR affect the sparsity of document-topic assignments? And what is the effect of re-estimated document-topic assignments on the topical diversity task?}


\section{Introduction}
\label{Introduction}
Quantitative notions of topical diversity in text documents are useful in several contexts, e.g., to assess the interdisciplinarity of a research proposal~\cite{Bache2013} or to determine the interestingness of a document~\cite{Azarbonyad2015}.
An influential formalization of diversity has been introduced in biology~\cite{Rao1982}. It decomposes diversity in terms of \emph{elements} that belong to \emph{categories} within a  \emph{population}~\cite{solow-measurement-1993} and formalizes the diversity of a population $d$  as the expected distance between two randomly selected elements of the population: 
\begin{equation}
div(d) = \sum_{i=1}^{T} \sum_{j=1}^{T} p_ip_j \delta(i,j),
\label{rao-div}
\end{equation}
where $p_i$ and $p_j$ are the proportions of categories $i$ and $j$ in the population and  $\delta(i, j)$ is the distance between  $i$ and $j$.  \citet{Bache2013} have adapted this notion of diversity to quantify the topical diversity of a text document. Words are considered elements, topics are categories, and a document is a population. When using topic modeling for measuring topical diversity of text document $d$, \citet{Bache2013} model elements based on the probability of a word $w$ given $d$, $P(w\conditional  d)$, categories based on the probability of $w$ given topic $t$, $P(w\conditional t)$, and populations based on the probability of $t$ given $d$, $P(t\conditional d)$.

In probabilistic topic modeling, at estimation time, these distributions are usually assumed to be sparse. First, the content of a document is assumed to be generated by a small subset of words from the vocabulary (i.e., $P(w\conditional  d)$ is sparse). Second, each topic is assumed to contain only some topic-specific related words (i.e., $P(w\conditional  t)$ is sparse). Finally, each document is assumed to deal with a few topics only (i.e., $(P(t\conditional  d)$ is sparse). 
When approximated using currently available methods, $P(w\conditional  t)$ and $P(t\conditional  d)$ are often dense rather than sparse \cite{Soleimani2015, Wallach2009, Lin2014}. Dense distributions cause two problems for the quality of topic models when used for measuring topical diversity: \emph{generality} and \emph{impurity}. General topics mostly contain general words and are typically assigned to most documents in a corpus.
Impure topics contain words that are not related to the topic.
Generality and impurity of topics both result in low quality $P(t\conditional  d)$ distributions.

We propose a hierarchical re-estimation process for making the distributions $P(w\conditional d)$, $P(w\conditional t)$ and $P(t\conditional d)$ more sparse. 
We re-estimate the parameters of these distributions so that general, collection-wide items are removed and only salient items are kept. 
For the re-estimation we use the concept of \textit{parsimony}~\cite{Hiemstra2004} to extract only essential parameters of each distribution.

Our main contributions are: \begin{inparaenum}[(1)]
\item We propose a hierarchical re-estimation process for topic models to address two main problems in estimating topical diversity of text documents, using a biologically inspired definition of diversity.
\item We study the efficacy of each level of re-estimation, and improve the accuracy of estimating topical diversity, outperforming the current state-of-the-art \cite{Bache2013} on a publicly available dataset commonly used for evaluating document diversity \cite{PubMed}.
\end{inparaenum}


\section{Related work}
\label{RelatedWork}
Our hierarchical re-estimation method for measuring topical diversity relates to measuring text diversity, improving the quality of topic models, model parsimonization, and evaluating topic models.

\textbf{Text diversity and interestingness.} 
Recent studies measure topical diversity of document~\cite{Bache2013, Derezinski2015, Azarbonyad2015} by means of Latent Dirichlet Allocation (LDA)~\cite{Blei2003}. 
The main diversity measure in this work is Rao's measure~\cite{Rao1982} (Equation \ref{rao-div}), in which the diversity of a text document is proportional to the number of dissimilar topics it covers.
While we also use Rao's measure, we hypothesize that pure LDA is not good enough for modeling text diversity and propose a re-estimation process for adapting topic models for measuring topical diversity. 

\textbf{Improving the quality of topic models.}
The two most important issues with topic models are the \emph{generality problem} and the \emph{impurity problem} \cite{Wallach2009, Soleimani2015, Boyd2014, Lin2014}.
Many approaches have been proposed to address the generality problem~\citep{Wallach2009, Wang2009, Williamson2010}.
The main difference with our work is that previous work does not yield sparse topic representations or topic word distributions. 
\citet{Soleimani2015} propose parsimonious topic models (PTM) to address the generality and impurity problems.
PTM achieves state-of-the-art results compared to existing topic models.
Unlike \cite{Soleimani2015}, we do not modify the training procedure of LDA but propose a method to refine the topic models.

\textbf{Model parsimonization. }
In language model parsimonization, the language model of a document is considered to be a mixture of a general background model and a document-specific language model \cite{Zhai2001, Hiemstra2004, Dehghani2016-ICTIR, Dehghani2016-CLEF}. 
The goal is to extract the document-specific part and remove the general words. 
We employ parsimonization for re-estimating topic models.
The main assumption in \cite{Hiemstra2004} is that the language model of a document is a mixture of its specific language model and a general language model:
\begin{equation}
\label{PLM-eq1}
P(w\conditional d) = \lambda P(w\conditional \tilde{\theta}_d) + (1-\lambda) P(w\conditional \theta_C),
\end{equation}
where $w$ is a term, $d$ a document, $\tilde{\theta}_d$ the document specific language model of $d$, $\theta_C$ the language model of the collection $C$, and $\lambda$ is a mixing parameter. The main goal is to estimate $P(w\conditional \tilde{\theta}_d)$ for each document.
This is done in an iterative manner using EM algorithm. The initial parameters of the language model are the parameters of standard language model, estimated using maximum likelihood: $P(w\conditional \tilde{\theta}_d) = \frac{\mathit{tf}_{w, d}}{\sum_{w'} \mathit{tf}_{w', d}}$,
where $\mathit{tf}_{w, d}$ is the frequency of $w$ in $d$. The following steps are computed iteratively:
\begin{description}
\item[\normalfont\emph{E-step}:] \mbox{}
\begin{equation}
\label{PLM-E-Step}
e_w = \mathit{tf}_{w, d} \cdot \frac{\lambda P(w\conditional \tilde{\theta}_{d})}{\lambda P(w\conditional \tilde{\theta}_{d}) + (1 - \lambda) P(w\conditional \theta_{C}))},
\end{equation}
\item[\normalfont\emph{M-step}:] \mbox{}
\begin{equation}
\label{eq3}
P(w\conditional \tilde{\theta}_{d}) = \frac{e_w}{\sum_{w'} e_{w'}},
\end{equation}
\end{description}
where $\tilde{\theta}_{d}$ is the parsimonized language model of document $d$, $C$ is the background collection, $P(w\conditional \theta_{C})$ is estimated using maximum likelihood estimation, and $\lambda$ is a parameter that controls the level of parsimonization. A low value of $\lambda$ will result in a more parsimonized model while $\lambda=1$ yields a model without parsimonization. 
The EM process stops after a fixed number of iterations or after convergence.

\textbf{Evaluating topic models.}
We evaluate the effectiveness of our re-estimated models by measuring the topical diversity of text documents.
In addition, in \S\ref{Analysis}, we analyze the effectiveness of our re-estimation approach in terms of purity in document clustering and document classification tasks. 
For classification, following \cite{Nguyen2015, Lacoste2009, Soleimani2015}, we model topics as document features with values $P(t\conditional d)$.
For clustering, each topic is considered a cluster and each document is assigned to its most probable topic \cite{Nguyen2015, Xie2013, Yan2013}.


\section{Measuring topical diversity of documents}
\label{MeasuringTopicalDiversityofDocuments}

To measure topical diversity of text documents, we propose \OurMethod{} (hierarchical topic model re-estimation).
\OurMethod{} can be applied to any topic modeling approach that models documents as distributions over topics and topics as distributions over words. 

\begin{figure}[h]
\centering
\includegraphics[height=1in]{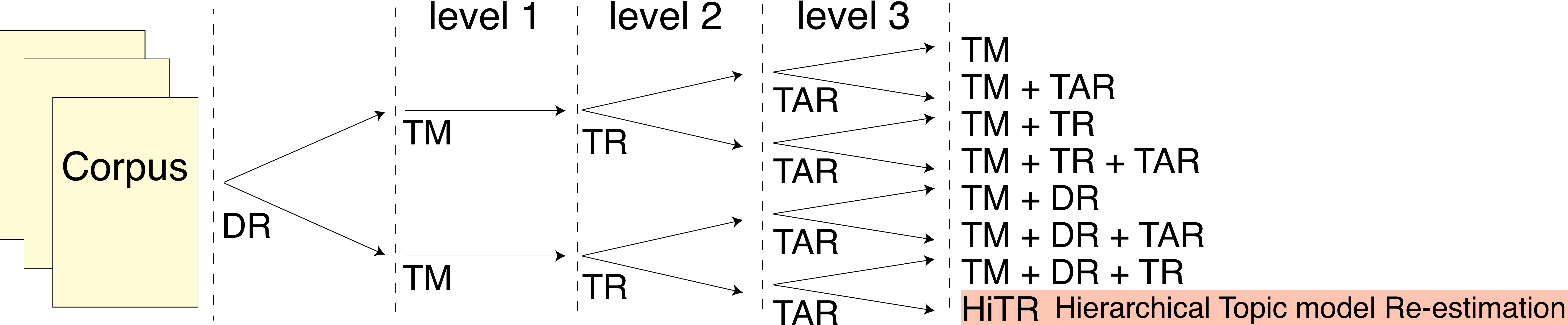}
\vspace*{0.1\baselineskip}
\caption{Different topic re-estimation approaches. TM is a topic modeling approach like, e.g., LDA. DR is document re-estimation, TR is topic re-estimation, and TAR is topic assignment re-estimation. 
\vspace{-0.5cm}}
\label{figure:PTM}
\end{figure}

\label{OurMethod}
The input to \OurMethod{} is a corpus of text documents.  
The output is a probability distribution over topics for each document in the corpus. 
\OurMethod{} has three levels of re-estimation:
(1)~\textbf{document re-estimation (DR)} re-estimates the language model per document $P(w\conditional  d)$;
(2)~\textbf{topic re-estimation (TR)} re-estimates the language model per topic $P(w\conditional  t)$; and 
(3)~\textbf{topic assignment re-estimation (TAR)} re-estimates the distribution over topics per document $P(t\conditional  d)$.
Based on applying or not applying re-estimation at different levels, there are seven possible re-estimation approaches;
see Fig.~\ref{figure:PTM}.
\OurMethod{} refers to the model that uses all three re-estimation techniques, i.e., TM+DR+TR+TAR. Next, we describe each of the re-estimation steps in more detail. 

%

\textbf{Document re-estimation (DR)}
\label{DP}
re-estimates $P(w\conditional  d)$.
Here, we remove unnecessary information from documents before training topic models. 
This is comparable to pre-processing steps, such as removing stopwords and high- and low-frequency words, that are typically carried out prior to applying topic models \cite{Blei2003, Nguyen2015,Mehrotra2013,Lau2014}. 
Proper pre-processing of documents, however, takes lots of effort and involves tuning many parameters.
\emph{Document re-estimation}, however, removes impure elements (general words) from documents automatically.
If general words are absent from documents, we expect that the trained topic models will not contain general topics. 
After document re-estimation, we can train any standard topic model on the re-estimated documents. 

Document re-estimation uses the parsimonization method described in \S\ref{RelatedWork}.
The re-estimated model $P(w\conditional  \tilde{\theta}_{d})$ in \eqref{eq3} is used as the language model of document $d$, and after removing unnecessary words from $d$, the frequencies of the remaining words (words with $P(w\conditional  \tilde{\theta}_{d}) > 0$) are re-estimated for $d$ using the following equation:
\begin{equation*}
\mathit{tf}(w, d) = \floor*{P(w\conditional  \tilde{\theta}_{d}) \cdot |d|},
\end{equation*}
where $|d|$ is the document length in words. Topic modeling is then applied on the re-estimated document-word frequency matrix.

\textbf{Topic re-estimation (TR)}
\label{TP}
re-estimates $P(w\conditional  t)$ by removing general words.
The re-estimated distributions are used to assign topics to documents.
The goal of this step is to increase the purity of topics by removing general words that have not yet been removed by DR.
The two main advantages of the increased purity of topics are\begin{inparaenum}[(1)]\item it improves human interpretation of topics, and \item it leads to more document-specific topic assignments, which is essential for measuring topical diversity of documents.\end{inparaenum} 

Our main assumption is that each topic's language model is a mixture of its topic-specific language model and the language model of the background collection.
TR extracts a topic-specific language model for each topic and removes the part that can be explained by the background model.
We initialize $\tilde{\theta}_{t}$ and $\theta_T$ as follows:
\[
P(w\conditional  \tilde{\theta}_{t}) = P(w\conditional  \theta_{t}^{\mathcal{TM}}) \qquad
P(w\conditional  \theta_T) = \frac{\sum_{t \in T} P(w\conditional  \theta_{t}^{\mathcal{TM}})}{\sum_{w' \in V} \sum_{t' \in T} P(w'\conditional  \theta_{t'}^{\mathcal{TM}})}
\]
where $t$ is a topic, $\tilde{\theta}_{t}$ is topic-specific language model of $t$, and $\theta_T$ is the background language model of $T$ (the collection of all topics), $P(w\conditional  \theta_{t}^{\mathcal{TM}})$ is the probability of $w$ belonging to topic $t$ estimated by a topic model $\mathcal{TM}$.
Having these estimations, the steps of TR are similar to the steps of parsimonization, except that in the E-step we estimate $\mathit{tf}_{w,t}$, the frequency of $w$ in $t$, by $P(w\conditional  \theta_{t}^{\mathcal{TM}})$.

\textbf{Topic assignment re-estimation (TAR)}
\label{TAP}
re-estimates $P(t\conditional  d)$.
In topic modeling, most topics are usually assigned with a non-zero probability to most of documents. For documents which are in reality about a few topics, this topic assignment is incorrect and overestimates its diversity.
TAR addresses the general topics problem and achieves more document specific topic assignments. 
To re-estimate topic assignments, a topic model is first trained on the document collection.
This model is used to assign topics to documents based on the proportion of words they have in common.
We then model the distribution over topics per document as a mixture of its document-specific topic distribution and the topic distribution of the entire collection.
%
%

We initialize $P(t\conditional  \tilde{\theta}_{d})$ and $P(t\conditional  \theta_C)$ as follows:
\[
 P(t\conditional  \tilde{\theta}_{d}) = P(t\conditional  \theta_d^{\mathcal{TM}}) \qquad
 P(t\conditional  \theta_C) = \frac{\sum_{d \in C} P(t\conditional  \theta_d^{\mathcal{TM}}) }{\sum_{t' \in T} \sum_{d' \in C} P(t'\conditional  \theta_{d'}^{\mathcal{TM}}) }.
\]
Here, $t$ is a topic, $d$ a document, $P(t\conditional  \tilde{\theta}_d)$ the document-specific topic distribution, and $P(t\conditional  \theta_C)$ the distribution of topics in the entire collection $C$, and $P(t\conditional  \theta_d^{\mathcal{TM}}) $ the probability of assigning topic $t$ to document $d$ estimated by a topic model $\mathcal{TM}$.
The remaining steps of TAR follow the ones of parsimonization, the difference being that in the E-step, we estimate $f_{t,d}$ using $P(t\conditional  \theta_d^{\mathcal{TM}})$.


\section{Experimental setup}
\label{ExperimentalSetup}

Our main research question is: (RQ1)~\rqone{}

To address RQ1 we run our models on a binary classification task.
We generate a synthetic dataset of documents with high and low topical diversity (the process is detailed below), and the task for every model is to predict whether a document belongs to the high or low diversity class. 
We employ \OurMethod{} to re-estimate topic models and use the re-estimated models for measuring topical diversity of documents.
To gain deeper insights into how \OurMethod{} performs, we conduct a separate analysis of the last two levels of re-estimation, TR and TAR:\footnote{As the DR level of re-estimation directly employs the parsimonious language modeling techniques in \cite{Hiemstra2004}, we omit it from our in-depth analysis.}
(RQ2.1)~\rqthree{} (RQ2.2)~\rqfour{}
To answer RQ2.1, we first evaluate the performance of TR on the topical diversity task and compare its performance to DR and TAR.
To answer RQ2.2, we first evaluate TAR together with LDA in a topical diversity task and analyze its effect on the performance of LDA to study how successful TAR is in removing general topics from documents. 

\textbf{Dataset, pre-processing, evaluation metrics, and parameters:} Following~\cite{Bache2013}, we generate 500 documents with a high value of diversity and 500 documents with a low value of diversity.
We select over 300,000 documents articles published between 2012 to 2015 from PubMed \cite{PubMed}.
For generating documents with a high value of diversity, we first select 20 journals and create 10 pairs of journals.
Each pair contains two journals that are relatively unrelated to each other (we use the pairs of journals selected in \cite{Bache2013}).
For each pair of journals $A$ and $B$ we select 50 articles to create 50 probability distributions over topics:
we randomly select one article from $A$ and one from $B$ and generate a document by averaging the selected article's bag of topic counts.
Thus, for each pair of journals we generate 50 documents with a high diversity value.
Also, for each of the chosen 20 journals,  we repeat the procedure but instead of choosing articles from different journals, we select them from the same journal to generate 25 non-diverse documents.

For pre-processing documents, we remove stopwords included in the standard stop word list from Python's NLTK package.
In addition, we remove the 100 most frequent words in the collection and words with fewer than 5 occurrences.

\textbf{Measuring topical diversity:} After re-estimating word distributions in documents, topics, and document topic distributions using \OurMethod{}, we use the final distributions over topics per document for measuring topical diversity.
Diversity of texts is computed using Rao's coefficient \cite{Bache2013} using Equation~\ref{rao-div}.
We use the normalized angular distance $\delta$ for measuring the distance between topics, since it is a proper distance function \cite{Azarbonyad2015}.

To measure the performance of topic models on the topical diversity task, we use ROC curves and report the AUC values~\cite{Bache2013}.
We also measure the \emph{coherence} of the extracted topics; this measure indicates the purity of $P(w\conditional t)$ distributions, where a high value of coherence implies high purity within topics. We estimate coherence using \emph{normalized pointwise mutual information} between the top $N$ words within a topic~\cite{Lau2014, Nguyen2015}.
As the reference corpus for computing word occurrences, we use the English Wikipedia.\footnote{We use a dump of June 2, 2015, containing 15.6 million articles.} 

The topic modeling approach used in our experiments with \OurMethod{} is LDA.
Following~\cite{Bache2013, Soleimani2015, Roder2015} we set the number of topics to 100.
We set the two hyperparameters to $\alpha=1/T$ and $\beta=0.01$, where $T$ is the number of topics, following~\cite{Nguyen2015}.
In the re-estimation process, at each step of the EM algorithm, we set the threshold for removing unnecessary components from the model to $0.0001$ and remove terms with an estimated probability less than this threshold from the language models, as in \cite{Hiemstra2004}.

We perform 10-fold cross validation, using 8 folds as training data, 1 fold to tune the parameters ($\lambda$ for DR, TR, and TAR), and 1 fold for testing.
Our baseline for the topical diversity task is the method proposed in \cite{Bache2013}, which uses LDA.
We also compare our results to PTM~\cite{Soleimani2015}, which we use instead of LDA for measuring topical diversity.
PTM is the best available topic modeling approach, and the current state of the art.

For statistical significance testing, we compare our methods to PTM using paired two-tailed t-tests with Bonferroni correction. 
To account for multiple testing, we consider an improvement significant  if: $p \leq \alpha/m$, where $m$ is the number of conducted comparisons and $\alpha$ is the desired significance. We set $\alpha=0.05$. 
In \S\ref{ExperimentalResults}, $^\blacktriangle$ and $^\blacktriangledown$ indicate that the corresponding method performs significantly better and worse than PTM, respectively.

\section{Results}
\label{ExperimentalResults}
In this section, we report on the performance of \OurMethod{} on the topical diversity task. 
Additionally we analyze the effectiveness of the individual re-estimation approaches.

\OurSection{Topical diversity results}

\begin{wrapfigure}[15]{r}{0.6\textwidth}
  \centering
  \includegraphics[width=0.58\textwidth]{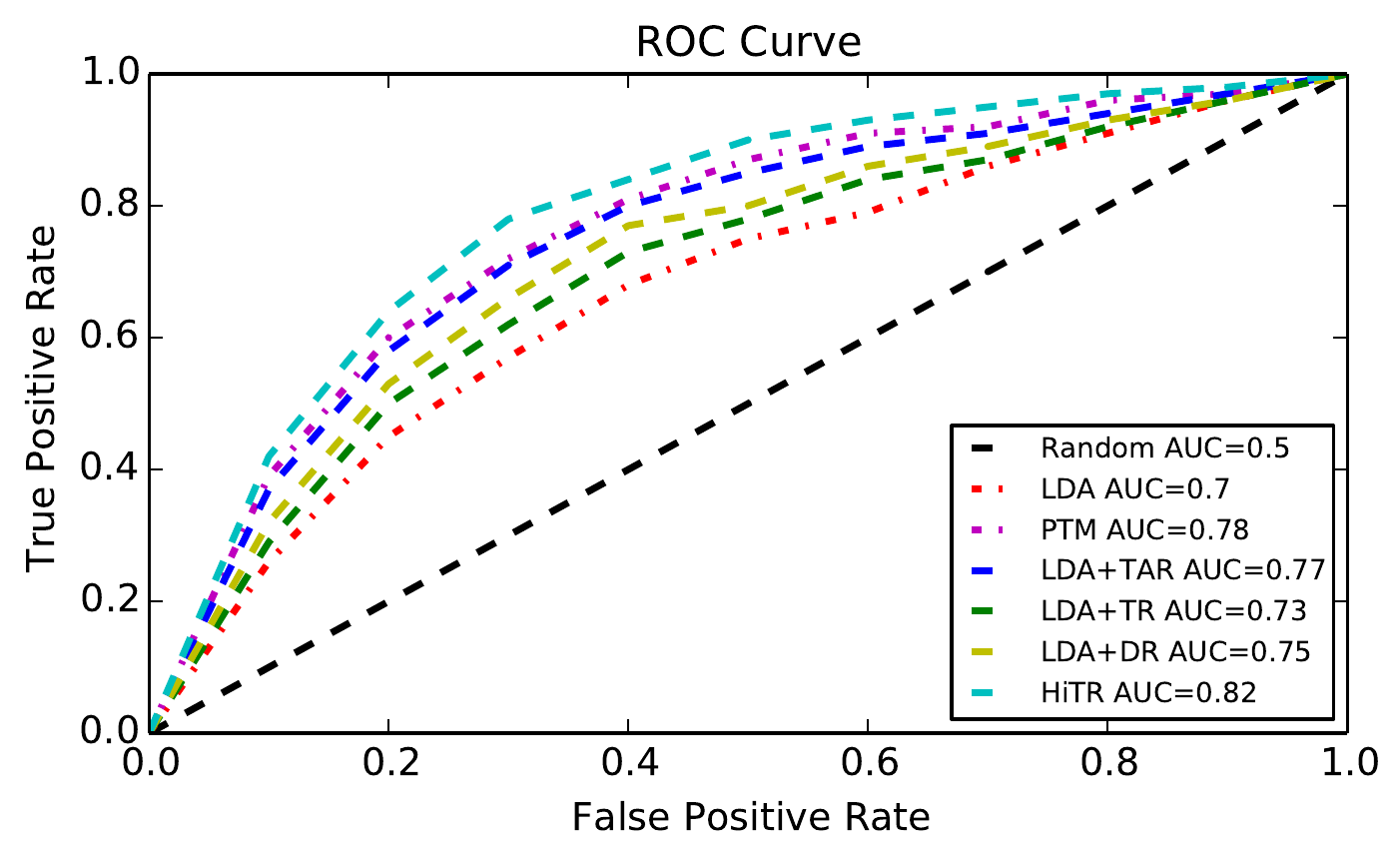}
  \caption{Performance of topic models in topical diversity task on the PubMed dataset. The improvement of \OurMethod{} over PTM is statistically significant ($p<0.05$) in terms of AUC.}
  \label{figure1}
\end{wrapfigure}

\label{threeLevelParsResults}
Fig.~\ref{figure1} plots the performance  of our topic models across different levels of re-estimation, and the models we compare to, on the Pub\-Med dataset. We plot ROC curves and compute AUC values. To plot the ROC curves we use the diversity scores calculated for the generated pseudo-documents with diversity labels.
\OurMethod{} improves the performance of LDA by 17\% and PTM by 5\% in terms of AUC. From Fig.~\ref{figure1} two observations can be made.
%
%
First, \OurMethod{} benefits from the three re-estimation approaches it encapsulates by successfully improving the quality of estimated diversity scores.
Second, the performance of LDA+TAR, which tries to address the generality problem, is higher than the performance of LDA+TR, which addresses impurity. General topics have a stronger negative effect on measuring topical diversity than impure topics. Also, LDA+DR outperforms LDA+TR. So, removing impurity from $P(t\conditional d)$ distributions is the most effective approach in the topical diversity task, and removing impurity from $P(w\conditional d)$ distributions is more effective than removing impurity from $P(w\conditional t)$ distributions.
Table \ref{table:divexample} illustrates the difference between LDA and \OurMethod{} with the topics assigned by the two methods for a non-diverse document that is combined from two documents from the same journal, entitled ``Molecular Neuroscience: Challenges Ahead'' and ``Reward Networks in the Brain as Captured by Connectivity Measures,'' using the procedure described in \S\ref{ExperimentalSetup}. As only a very basic stopword list being applied, words like \textit{also} and \textit{one} still appear. We expect to have a low diversity value for 
%
%
\begin{table}[b]
  \centering 
  \vspace{-2\baselineskip}
  \caption{\label{table:divexample}
    Topic assignments for a non-diverse document using LDA and \OurMethod{}. Only topics with $P(t\conditional d) > 0.05$ are shown.} 
  \resizebox{\textwidth}{!}{  
    \begin{tabular}{c c l c l}
      \toprule
      \multicolumn{3}{c}{LDA} & \multicolumn{2}{c}{\OurMethod{}}\\
      Topic & $P(t\conditional  d)$ & Top 5 words & $P(t\conditional  d)$ & Top 5 words \\
      \midrule
      1 & 0.21 & brain, anterior, neurons, cortex, neuronal & 0.68 & brain, neuronal, neurons, neurological,  nerve \\
      2 & 0.14 & channel, neuron, membrane, receptor, current & 0.23 & channel, synaptic, neuron, receptor, membrane\\
      3 & 0.10 & use, information, also, new, one & 0.09 & network, nodes, cluster, community, interaction\\
      4 & 0.08 & network, nodes, cluster, functional, node \\
      5 & 0.08 & using, method, used, image, algorithm & &\\
      6 & 0.08 & time, study, days, period, baseline & &\\
      7 & 0.07 & data, values, number, average, used & &\\
      \bottomrule
    \end{tabular}
  }
  \vspace{-0.5cm}
\end{table}
%
%
the combined document. However, using Rao's  diversity measure, the topical 
diversity of this document based on the LDA topics is 0.97.
This is due to the fact that there are three document-specific topics---topics 1, 2 and 4---and four general topics. Topics~1 
and~2 are very similar and their $\delta$ is 0.13. The other, more general topics have high $\delta$ values; the average $\delta$ value between pairs of topics is as high as 0.38.
For the same document, \OurMethod{} only assigns three document-specific topics and they are more pure and coherent. 
The average $\delta$ value between pairs of topics assigned by \OurMethod{} is 0.19. The diversity value of this document using \OurMethod{} is 0.16, which indicates that this document is non-diverse. 
Hence, \OurMethod{} is more effective than other approaches in measuring topical diversity of documents; it
successfully removes generality from $P(t\conditional  d)$.

\OurSection{Topic re-estimation results}
\label{topicParsimonizationResults}

To answer \textbf{RQ2.1}, we focus on topic re-estimation (TR). Since TR tries to remove impurity from topics, we expect it to increase the coherence of the topics by removing unnecessary words from topics. 
We measure the purity of topics based on the coherence of words in $P(w\conditional t)$ distributions. Table~\ref{table:5} shows the coherence of topics according to different topic modeling approaches, in terms of average mutual information. 
TR significantly increases the coherence of topics by removing the impure parts from topics. The coherence of PTM is higher than of TR. However, when we first apply DR, train LDA, and finally apply TR, the coherence of the extracted topics is significantly higher than the coherence of topics extracted by PTM.
We conclude that TR is effective in removing impurity from topics. Moreover, DR also contributes in making topics more pure.

\begin{table}[t]
	\centering
	\caption{Topic model coherence in terms of average normalized mutual information between top 10 words in the topics on the PubMed dataset.} 
	\label{table:5}
	\begin{tabularx}{0.7\textwidth}{CCCC}
          \toprule	
          \textbf{LDA} & \textbf{PTM} & \textbf{LDA+TR} & \textbf{LDA+DR+TR} \\
          \midrule 
          8.17 & 9.89 & 9.46 & 10.29$^\blacktriangle$ \\    
          \bottomrule
	\end{tabularx}
	 \vspace*{-1.5\baselineskip}
\end{table}

\OurSection{Topic assignment re-estimation results}
\label{topicAssignmentParsimonizationResults}

To answer \textbf{RQ2.2}, we focus on TAR (topic assignment re-estimation). We are interested in seeing how \OurMethod{} deals with general topics. We sum the probability of assigning a topic to a document, over all documents: 
\begin{wrapfigure}[18]{r}{0.6\textwidth}
  \centering
  \vspace*{-1.5\baselineskip}
  \includegraphics[width=0.58\columnwidth]{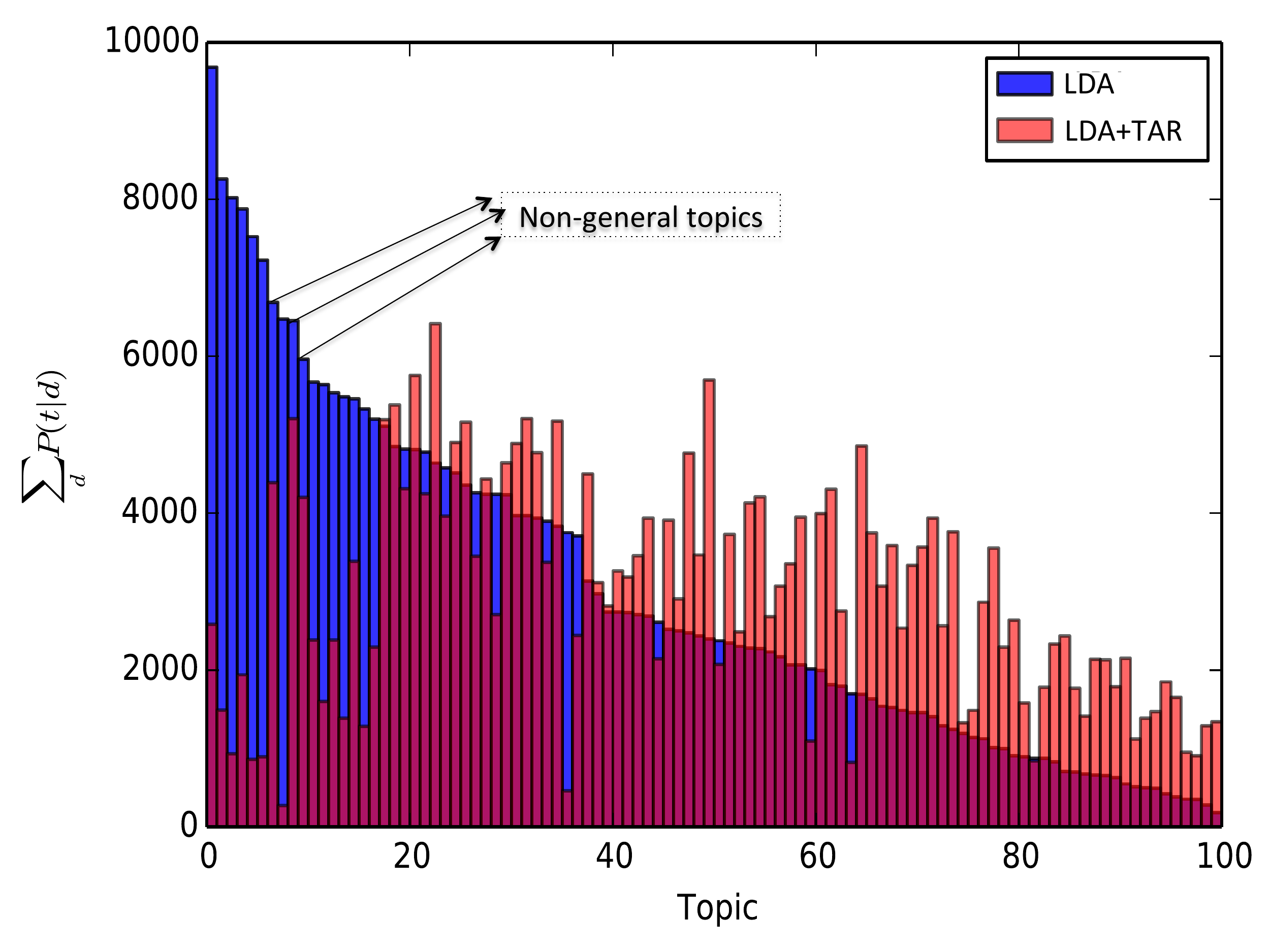}
  \caption{The total probability of assigning topics to the documents in the PubMed dataset estimated using LDA and LDA+TAR. (The two areas are equal to the number of documents ($N\approx300K$)).}
\label{figure4}
\end{wrapfigure}
%
%
for each topic $t$, we compute $\sum_{d \in C} P(t\conditional d)$, where $C$ is the document collection. Fig.~\ref{figure4} shows the distribution of probability mass before and after applying TAR;
topics are sorted based on the topic assignment probability of LDA.  LDA assigns a vast proportion of the probability mass to a relatively small number of topics, mostly general topics that are assigned to most documents.  
We expect that many topics are represented in some documents, while relatively few topics will be relevant to all documents. After applying TAR, the distribution is less skewed and the probability mass is more evenly distributed.

There are topics that have a high $\sum_d P(t\conditional d)$ value in LDA's topic assignments and a high $\sum_d P(t\conditional d)$ value after applying TAR too; we marked them as ``non-general topics'' in Fig.~\ref{figure4}.
Table~\ref{table:6}, column~2 shows the top five words for these topics. 
TAR is able to find these three non-general topics and their assignment probabilities to documents in the $P(t\conditional d)$ distributions is not changed 
as much as the actual general topics. 
Thus, TAR removes general topics from documents and increases the probability of document-specific topics for each document.
To further investigate whether TAR really removes general topics, Table~\ref{table:6}, column~3 shows the top five words for the first 10 topics in Fig.~\ref{figure4}, excluding the ``non-general topics.''
These seven topics have the highest decrease in  $\sum_d P(t\conditional d)$ values due to TAR.  Clearly, they contain general words and are not informative.
Fig.~\ref{figure4} shows that after applying TAR, the $\sum_d P(t\conditional d)$ values have decreased dramatically for these topics, without creating new general topics.

\begin{table}[t]
  \centering 
  \caption{\label{table:6}
    Top five words for the topics detected by TAR as general topics and non-general topics.} 
  \resizebox{\textwidth}{!}{  
    \begin{tabular}{c l l}
      \toprule
      Topic & Non-general topics & General topics \\
      \midrule
      1
      & health, services, public, countries, data
      & use, information, also, new, one \\
      2
      & surgery, surgical, postoperative, patient, performed
      & ci, study, analysis, data, variables\\
      3
      & cells, cell, treatment, experiments, used
      & time, study, days, period, baseline \\
      4
      &
      & group, control, significantly, compared, groups \\
      5
      &
      & study, group, subject, groups, significant \\
      6
      &
      & may, also, effects, however, would \\
      7
      &
      & data, values, number, average, used \\
      \bottomrule
    \end{tabular}
  }
  \vspace{-1.5\baselineskip}
\end{table}

\begin{wrapfigure}[13]{r}{0.6\textwidth}
  \centering
  \vspace*{-1.5\baselineskip}
\includegraphics[width=0.58\columnwidth]{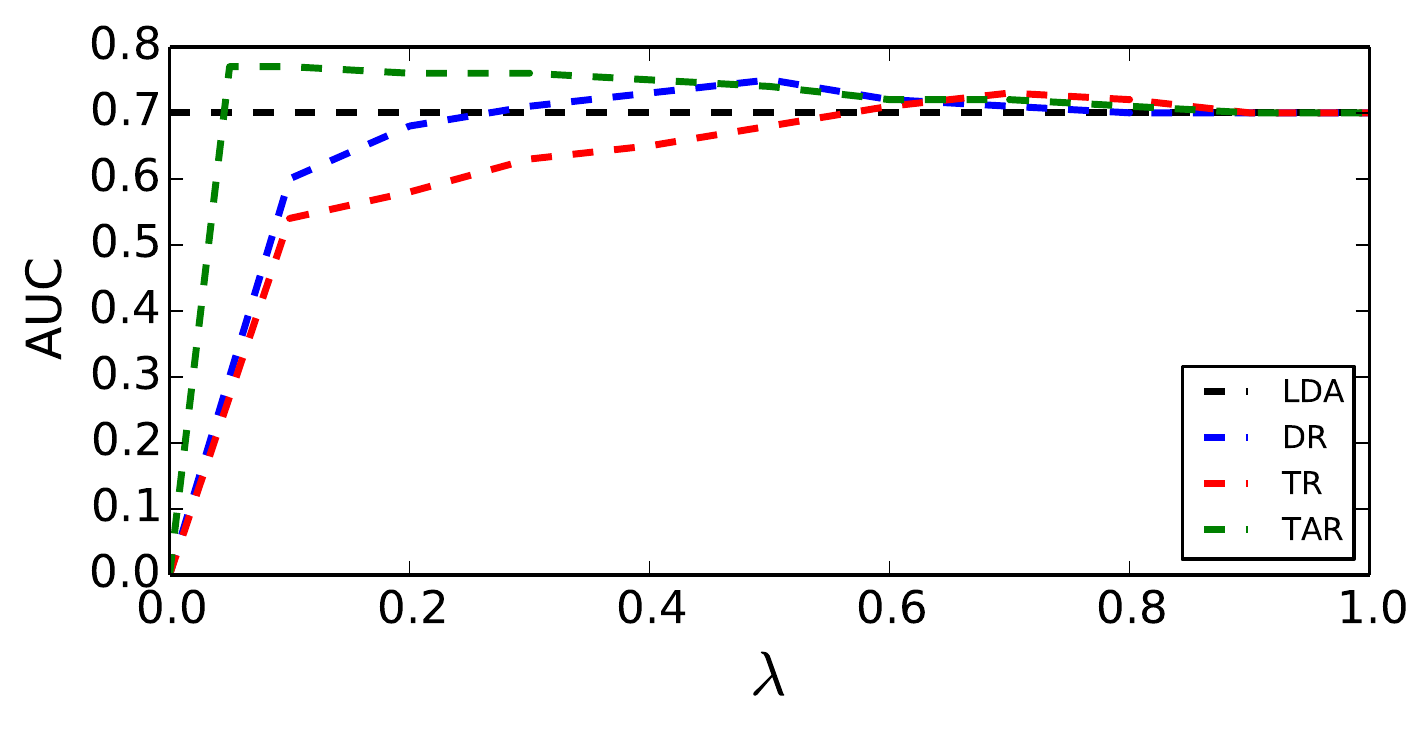}
  \caption{The effect of the $\lambda$ parameter on the performance of topics models in the topical diversity task on the PubMed dataset.}
\label{figure2}
\end{wrapfigure}

\OurSection{Parameter analysis}
Next, we analyze the effect of the $\lambda$ parameter on the performance of DR, TR, and TAR.
Fig.~\ref{figure2} displays the performance at different levels of re-estimation. With $\lambda=1$, no re-estimation  occurs, and all methods equal LDA.
%
We see that DR peaks with moderate values of $\lambda$ ($0.4 \leq \lambda \leq 0.45$). This reflects that documents contain a moderate amount of general information and that DR is able to successfully deal with it. For $\lambda \geq 0.8$, the performance of DR and LDA is the same and for these values of $\lambda$ DR does not increase the quality of LDA. Also, the best performance of TR is achieved with high values of $\lambda$ ($0.65 \leq \lambda \leq 0.75$).
From this observation we conclude that topics typically need only a small amount of re-estimation. With this slight re-estimation, TR is able to improve the quality of LDA. However, for $\lambda \geq 0.75$ the accuracy of TR degrades.
Lastly, TAR achieves its best performance with low values of $\lambda$ ($0.02 \leq \lambda \leq 0.05$). Hence, most of the noise is in the $P(t\conditional d)$ distributions and aggressive re-estimation allows TAR to remove most of it. 


\section{Analysis}
\label{Analysis}
In this section, we want to gain additional insights into \OurMethod{} and its effects on topic computation.
The purity of topic assignments based on $P(t\conditional d)$ distributions has the highest effect on the quality of estimated diversity scores. Thus, we investigate how pure estimated topic assignments are using \OurMethod{}. 
To this end, we compare document clustering and classification results, based on the topics assigned by \OurMethod{}, LDA and PTM.
For clustering, following~\cite{Nguyen2015}, we consider each topic as a cluster. Each document $d$ is assigned to the topic that has the highest probability value in $P(t\conditional d)$.
For classification, we use all topics assigned to the document and consider $P(t\conditional d)$ as features for a supervised classification algorithm; we use SVM. 
We view high accuracy in clustering or classification as an indicator of high purity of topic distributions. 
Our focus is not on achieving a top clustering or classification performance: these tasks are a means to assess the purity of topic distributions using different topic models.

\textbf{Datasets and metrics.}
We use RCV1~\cite{Lewis2004}, 20-NewsGroups,\footnote{Available at \small{\url{http://www.ai.mit.edu/people/~jrennie/20Newsgroups/}}} and Ohsu\-med.\footnote{Available at \small{\url{http://disi.unitn.it/moschitti/corpora.htm}}}
RCV1 contains 806,791 documents with category labels for 126 categories. For clustering and classification of documents, we use 55 categories in the second level of the hierarchy. 20-NewsGroups contains ${\sim}$20,000 documents (20 categories, around 1,000 documents per category). Ohsumed contains 50,216 documents grouped into 23 categories.
For measuring the purity of clusters we use \emph{purity} and \emph{normalized mutual information} (NMI) \cite{Manning2008}.
We use 10-fold cross validation and the same pre-processing as in \S\ref{ExperimentalSetup}.

\textbf{Purity results.}
The top part of Table~\ref{table:1} shows results on the document clustering task.
As we can see, the topic distributions extracted using \OurMethod{} score higher than the ones extracted using LDA and PTM in terms of both purity and NMI. This shows the ability of \OurMethod{} to make $P(t\conditional  d)$ more pure. 
The two-level re-estimated topic models achieve higher purity values than their respective one-level counterparts except the combination of DR and TR, which indicates that re-estimation at each level contributes to the purity of $P(t\conditional  d)$. 
The combination of TR and DR is not effective in increasing purity over its one-level counterparts on most of the datasets, indicating that TR and DR address similar issues. But when each of them is combined with TAR, the purity of the topic distributions increases, implying that DR/TR and TAR address complementary issues.

\begin{table*}[t]
	\centering 
	\caption{\label{table:1}
Re-estimated topic models for document clustering (top) and document classification (bottom). For significance tests, we consider {p-value} $<$ 0.05/7; comparisons are against PTM. 
}
\begin{tabular}{l c c c c c c}
    \toprule
    & \multicolumn{2}{c}{\textbf{RCV1}} & \multicolumn{2}{c}{\textbf{20-Newsgroups}} & \multicolumn{2}{c}{\textbf{Ohsumed }} \\
   \textbf{Method} & Purity & NMI & Purity & NMI & Purity & NMI\\
   \midrule
   LDA & 0.55~~ & 0.40~~ & 0.52~~ & 0.36~~ & 0.50~~ & 0.30~~ \\
   PTM & 0.61~~ & 0.43~~ & 0.57~~ & 0.38~~ & 0.55~~ & 0.33~~ \\
   \midrule
   LDA+DR & 0.57$^\blacktriangledown$ & 0.41$^\blacktriangledown$ & 0.56~~ & 0.39~~ & 0.53$^\blacktriangledown$ & 0.32$^\blacktriangledown$ \\
   LDA+TR & 0.57$^\blacktriangledown$ & 0.42$^\blacktriangledown$ & 0.56~~ & 0.38~~ & 0.53$^\blacktriangledown$ & 0.31$^\blacktriangledown$ \\
   LDA+TAR & 0.60~~ & 0.43~~ & 0.57~~ & 0.39~~ & 0.54~~ & 0.33~~ \\
   LDA+DR+TR & 0.58~~ & 0.42$^\blacktriangledown$ & 0.57~~ & 0.38~~ & 0.54~~ & 0.32~~ \\
   LDA+DR+TAR & 0.60~~ & 0.43~~ & 0.58~~ & 0.40~~ & 0.55~~ & 0.35$^\blacktriangle$ \\
   LDA+TR+TAR & 0.61~~ & 0.43~~ & 0.58~~ & 0.40$^\blacktriangle$ & 0.56$^\blacktriangle$ & 0.34$^\blacktriangle$ \\
   \midrule
   \OurMethod{} & \textbf{0.64}$^\blacktriangle$  &  \textbf{0.45}$^\blacktriangle$  & \textbf{0.60}$^\blacktriangle$  &  \textbf{0.42}$^\blacktriangle$  & \textbf{0.57}$^\blacktriangle$  &  \textbf{0.35}$^\blacktriangle$ \\
   \bottomrule
%
\label{table:2}
            & Acc.  & Change & Acc. & Change   & Acc. & Change  \\
            \midrule
            LDA & 0.76~~ & \phantom{0}-8\% & 0.81~~ & \phantom{0}-7\% & 0.50~~ & \phantom{0}-11\%\\
            PTM & 0.82~~ & -- & 0.87~~ & -- & 0.56~~ & -- \\
            \midrule
            LDA+DR & 0.79$^\blacktriangledown$ &  & 0.83$^\blacktriangledown$ & \phantom{0}-5\% & 0.52$^\blacktriangledown$ & \phantom{0}-7\% \\
            LDA+TR & 0.78$^\blacktriangledown$ & \phantom{0}-5\% & 0.83$^\blacktriangledown$ & \phantom{0}-5\% & 0.53$^\blacktriangledown$ & \phantom{0}-5\% \\
            LDA+TAR & 0.82~~ & \phantom{0}0\% & 0.85$^\blacktriangledown$ & \phantom{0}-2\% & 0.54~~ & \phantom{0}-4\% \\
            LDA+DR+TR & 0.80$^\blacktriangledown$ & \phantom{0}-2\% & 0.84$^\blacktriangledown$ & \phantom{0}-3\% & 0.53$^\blacktriangledown$ & \phantom{0}-5\% \\
            LDA+DR+TAR & 0.83~~ & \phantom{0}+1\% & 0.86~~ & \phantom{0}-1\% & 0.56~~ & \phantom{0}0\% \\
            LDA+TR+TAR & 0.82$^\blacktriangle$ & \phantom{0}0\% & 0.87~~ & \phantom{0}0\% & 0.58$^\blacktriangle$ & \phantom{0}+4\% \\
            \midrule
            \OurMethod{}  &  \textbf{0.85}$^\blacktriangle$ & \phantom{0}+4\% &  \textbf{0.89}$^\blacktriangle$ & \phantom{0}+2\% &  \textbf{0.60}$^\blacktriangle$ & \phantom{0}+7\% \\
	    \bottomrule
	  \end{tabular}
\vspace*{-2\baselineskip}
\end{table*}	

The bottom part of Table~\ref{table:2} shows results on the document classification task.
\OurMethod{} is more accurate in estimating $P(t\conditional  d)$; its accuracy is higher than that of other topic models.
The higher values in classification task, compared to clustering task, indicate that the most probable topic does not necessarily contain all information about the content of a document.
If a document is about more than one topic, the classifier utilizes all $P(t\conditional d)$ information and performs better. Therefore, the higher accuracy of \OurMethod{} in this task is an indicator of its ability to assign document-specific topics to documents. 

\section{Conclusions}
\label{conclusion}
We have proposed Hierarchical Topic model Re-estimation (\OurMethod{}), an approach for measuring topical diversity of text documents.
It addresses two main issues with topic models, topic generality and topic impurity, which negatively affect measuring topical diversity scores in three ways.
First, the existence of document-unspecific words within $P(w\conditional d)$ (the distribution of words within documents) yields general topics and impure topics.
Second, the existence of topic-unspecific words within $P(w\conditional t)$ (the distribution of words within topics) yields impure topics.
Third, the existence of document-unspecific topics within $P(t\conditional d)$ (the distribution of topics within documents) yields general topics.
We have proposed three approaches for removing unnecessary or even harmful information from  probability distributions, which we combine in our method for \OurMethod{}.

Estimated diversity scores for documents using \OurMethod{} are more accurate than those obtained using the current state-of-the-art topic modeling method PTM, or a general purpose topic model such as LDA. \OurMethod{} outperforms PTM because it adapts topic models for the topical diversity task.
The quality of topic models for measuring topical diversity degrades mainly because of general topics in the $P(t\conditional d)$ distributions. 
Our topic assignment re-estimation (TAR) approach successfully removes general topics, leading to higher performance on the topical diversity task.

 We analyzed the purity of topic assignments on clustering and classification tasks, where $P(t\conditional d)$ distributions were directly used as features.
The results confirm that \OurMethod{} is effective in removing impurity from documents; it removes impure parts from the three probability distributions mentioned, using three re-estimation approaches.

\smallskip
\begin{spacing}{1}
\noindent\small
\textbf{Acknowledgments.}
This research was supported by
Ahold Delhaize,
Amsterdam Data Science,
Blendle,
the Bloomberg Research Grant program,
the Dutch national program COMMIT,
Elsevier,
the European Community's Seventh Framework Programme (FP7/2007-2013) under
grant agreements nr 283465 (ENVRI) and 312827 (VOX-Pol),
the Microsoft Research Ph.D.\ program,
the Netherlands eScience Center under project number 027.012.105,
the Netherlands Institute for Sound and Vision,
the Netherlands Organisation for Scientific Research (NWO)
under pro\-ject nrs
314.99.108,
600.006.014,
HOR-11-10, 
CI-14-25, 
652.\-002.\-001, 
612.\-001.\-551, 
652.\-001.\-003, 
314-98-071, 
and
Yandex.
All content represents the opinion of the authors, which is not necessarily shared or endorsed by their respective employers and/or sponsors.
\end{spacing}

\renewcommand{\bibsection}{\section{References}}
\bibliographystyle{abbrvnat}
\setlength{\bibsep}{0pt}
{
\small
  \bibliography{ref}
}

\end{document}